**Metal Enhanced Interactions of Graphene with Monosaccharides**

A Manuscript Submitted for publication to

*Chemical Physics Letters*

February 15, 2016

Carlos Pereyda-Pierre[a] and Abraham F. Jalbout[b*]

[a]DIFUS, Universidad de Sonora, Hermosillo, Sonora, México

[b]Centro de Investigación en Química Aplicada, Saltillo, Coahuila, México

*Corresponding Author
E-mail address: drajalbout@gmail.com

ABSTRACT

The present theoretical study proposes the enhanced interaction of nanostructures with monosaccharides. It has been demonstrated that the interactions with and without metal adsorption do in fact show that the adsorption energies change accordingly. It is important to note that the chemistry of reactions can also be influenced as a result of this change in dynamics.

INTRODUCTION

Graphene has been referenced a lot in recent literature whereby numerous theoretical studies [1-2]. It has been demonstrated that in the $Li^+$ case, a charge of ~0.8 $e$ is transferred to the surface of the fullerene ($C_{60}$)molecule [3] and small graphene sheets [1]. Modification of the surface electron configuration was also shown to yield increased reactivity. There has been many similar studies on nanotubes [4-6] whereby the adsorption of metal ions can influence chemistry of the graphene based system.

We shall find as central result of this letter that metal adsorption has an effect on the ability of sugars to adsorb to the surface. This effect can be explained in terms of a qualitative discussion of bond stretching while results on quasi 1-D molecules such as polyacenes and polyphenyls showing similar distortions may indicate some quantum-phase transition, which is hidden by the fact that adsorption happens in finite steps.

In previous studies [7] the interaction between various carbohydrates and a fragment of zigzag (10,0) CNT has been shown to be favourable and exothermic in the range of -6 to -8 kcal/mol. What is important to gain from our analysis is that the interaction energies do change upon adsorption of the metal ions that helps to reinforce the concept that this adsorption will play a key role for surface chemistry.

COMPUTATIONAL METHODS

The quantum chemical calculations performed in this work on all structures were done by closing the sheets with hydrogen atoms as we assume finite conditions. The GAUSSIAN03[8]codes were used. Since the systems are relatively large the geometry optimizations were performed with the Density functional theory (DFT) B3LYP method. The basis set used is the 6-31+G* for optimizations and single point energy evaluations. All calculations were verified to ensure stability of the systems obtained.

## RESULTS AND DISCUSSION

All optimized structures for the calculations are presented (FIGURES 1-3) wherebyselected geometrical parameters shown for optimized structures without Li (A) and with Li (B) whereby angles are shown in angles (°) and bond distances in angstroms (Å). The structures are denoted as glucose (**1**), fructose (**2**), galactose (**3**), mannose (**4**) and ribose (**5**). And in the table (TABLE 1) relative dissociation energies in kcal/mol where a. is without metal adsorption and b. is with metal adsorption. The energies shown are at the B3LYP/6-31+G* level of theory with optimizations of the stable structures.

From the dissociation energies we can observe that for glucose without the metal we obtain a value of 4.10 kcal/mol which if we look at the result of the system with Li it is higher at 4.66 kcal/mol and actually endothermic. This means that it actually requires less energy to dissociate relative to that without the metal. The electronic transfer properties of the Li are maintained which is encouraging.

For next system, fructose, the dissociation energy without the metal is -4.07 kcal/mol that is within error of the calculations themselves. This essentially means that the systems would in fact repel each other. After exposure to the metal we can see an increase to 4.45 kcal/mol which is similar to that seen for glucose that is logical given the similarity in molecular structure.

This trend is again observed in galactose with an important dissociation energy of -4.34 kcal/mol when it is included the metal, and the corresponding 3.61 kcal/mol when we consider the system without the presence of the metal.

A different behaviour to glucose, fructose and galactose is distinguished in the respective dissociation energies for the mannose system: values of 4.97 kcal/mol without metal and 4.93 kcal/mol with metal, in other words, we appreciate a dissociation energy decrease in comparison to the presence of metal and the absence of it.

In the last system with ribose we see again a significant increase in the dissociation energies after metal exposure; 5.28 kcal/mol with metal and 2.59 kcal/mol without metal.

The general trend that we can observe in the results is that upon exposure to a metal ion the electronic affinity of the system (due to charge transfer and symmetry breaking) increases the interaction. This enhances our assumption that metal ions can make systems more susceptible to interactions with organic compounds.

## CONCLUSIONS

The results are very interesting and present a detailed description of the reaction dynamics involved adsorption between graphene flakes and monosaccharides of different sizes. The tendencies of the structures to undergo further interactions as a result of metal adsorption are observed by the presented calculations. To the best of our knowledge no similar study has been presented and helps to validate the changing dynamics of sugar adsorption to graphene or carbon molecular surfaces after interaction by metal ions. If these metal ions can be used to manipulate surface chemistry it can lead to interesting dynamics for surface catalysis, molecule manufacturing and new chemical process design.

The main conclusions to extract from this work are that upon interaction of the metal ions with the molecular surface charge transfer permits the surfaces to become much more reactive. This can lead us to believe that if we want to use the surface for catalysis or formation of sugar components the metals play a vital role to this effect. This is important and is a trend that has been observed in other related systems when using metal ions [1-6].

In the case of Ribose for example we have seen that the size of the sugar has a less profound effect on whether metals are used or not. As we can see the energies with and without the metal are quite similar. It is important for small molecule catalysis, however we will verify this later.

In a further step experimental validation can be performed to determine whether our proposed solution of chemical modification is indeed possible by interaction of a metal source with graphene mixtures. We will pursue further theoretical evidence revealing interesting dynamics along this nature to help build the foundation for future validation by experiment. If we can pursue the relevant information to other applications such as waste management [9] these metal adsorption methods can be important for chemical and physical surface science.

When increasing the metal adsorption to two Li atoms to the flake (see FIGURE 3) we observe interesting geometrical distortions. The dissociation energy of the $Li_2$@Flake-Glucose system into $Li_2$@Flake and Glucose is 5.45 kcal/mol representing an increase of 17% with the single metal (Li@Flake-Glucose) system. This is important since it is clear that we can use metal adsorption to control the affinity of the monosaccharides to the graphene surface. Also, we see local adsorption of the electron density from the Li atoms leading to symmetry breaking and increased dissociation energy.

It is interesting to note that the geometrical distortions upon the metal adsorption in the above mentioned systems whereby electron repulsion causes the separation of the monosaccharides to the surface. This is especially true for the ribose system where increased steric hindrance and rotation causes the distances and distortion to increase even further as a result of the OH groups. Without the metal adsorption the monosaccharides are localized on the centre of the flake. Once the metal is adsorbed we observe the symmetry breaking and distortion of the flake. The dissociation energies represent an interesting tendency upon metal adsorption allowing control of the interaction process by use of metal ions.

FIGURE AND TABLE CAPTIONS

FIGURE 1. Selected geometrical parameters optimized at the B3LYP/6-31+G* level of theory shown for the optimized structures without Li whereby angles are shown in angles (∘) and bond distances in angstroms (Å). Where structures are denoted as glucose (**1**), fructose (**2**), galactose (**3**), mannose (**4**) and ribose (**5**).

FIGURE 2. Selected geometrical parameters optimized at the B3LYP/6-31+G* level of theory shown for the optimized structures with Li whereby angles are shown in angles (∘) and bond distances in angstroms (Å). Where structures are denoted as glucose (**1**), fructose (**2**), galactose (**3**), mannose (**4**) and ribose (**5**).

FIGURE 3. Selected geometrical parameters optimized at the B3LYP/6-31+G* level of theory shown for the optimized structures with Li whereby angles are shown in angles (∘) and bond distances in angstroms (Å) for the system of $Li_2$@Flake-Glucose.

TABLE 1. Relative dissociation energies in kcal/mol where **a**. is *without metal adsorption* and **b.** is *with metal adsorption*. Please see description from FIGURE 1 for structural designations. They are calculated at the B3LYP/6-31+G* level of theory.

ACKNOWLEDGEMENTS

We appreciate support from CONACyT, CIQA in the use of installations and resources and ACARUS in the use of its computational resources.

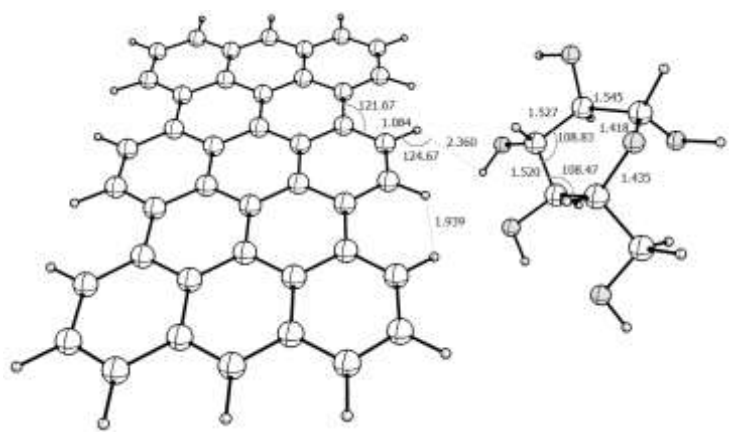

**1**

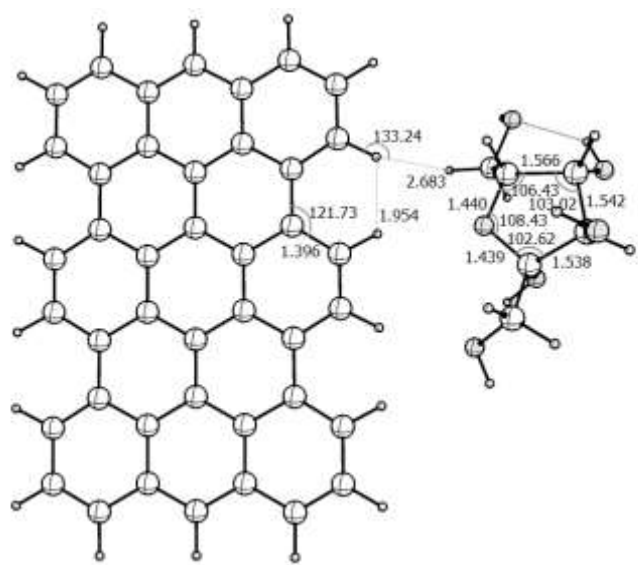

**2**

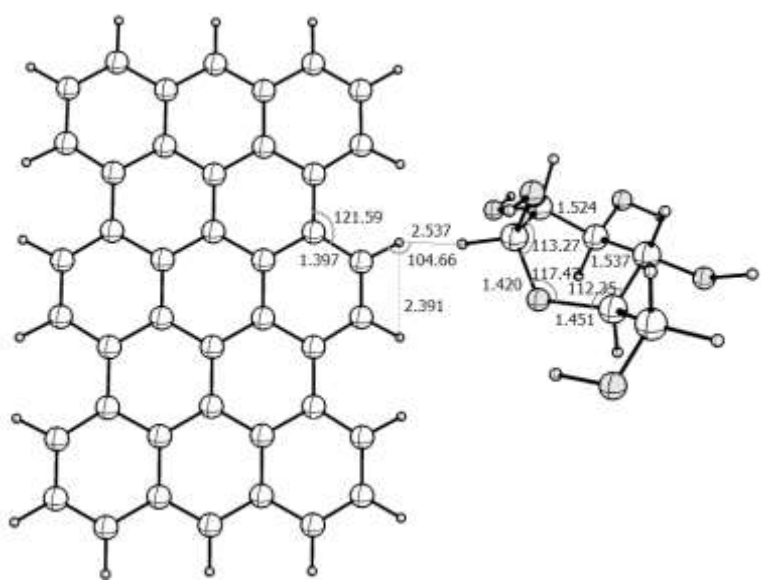

**3**

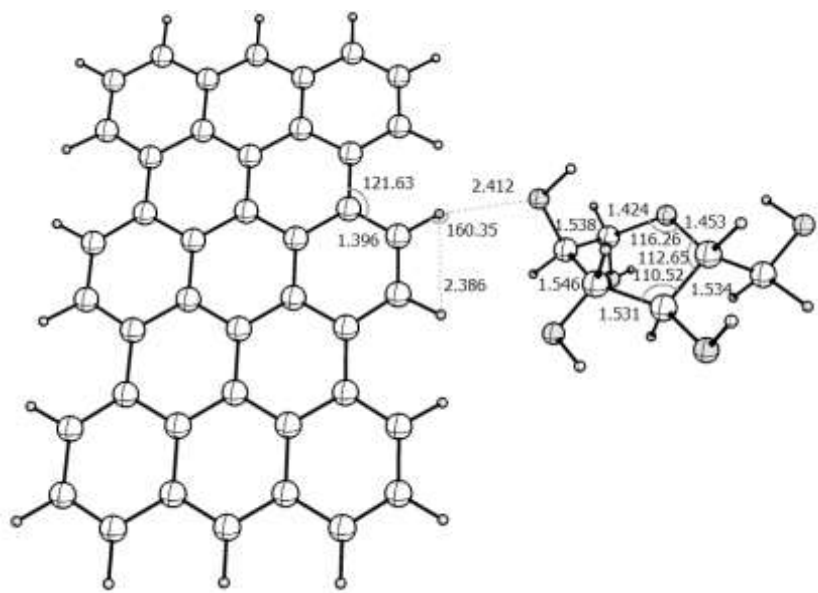

**4**

**5**

FIGURE 1.

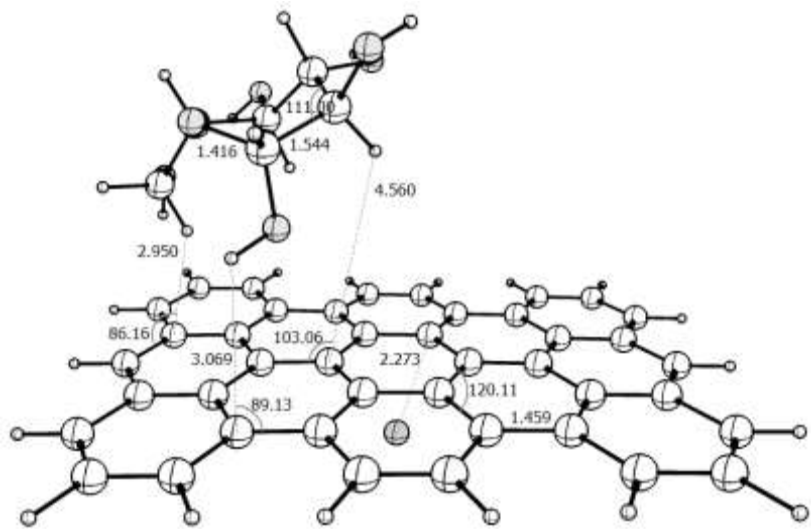

**1**

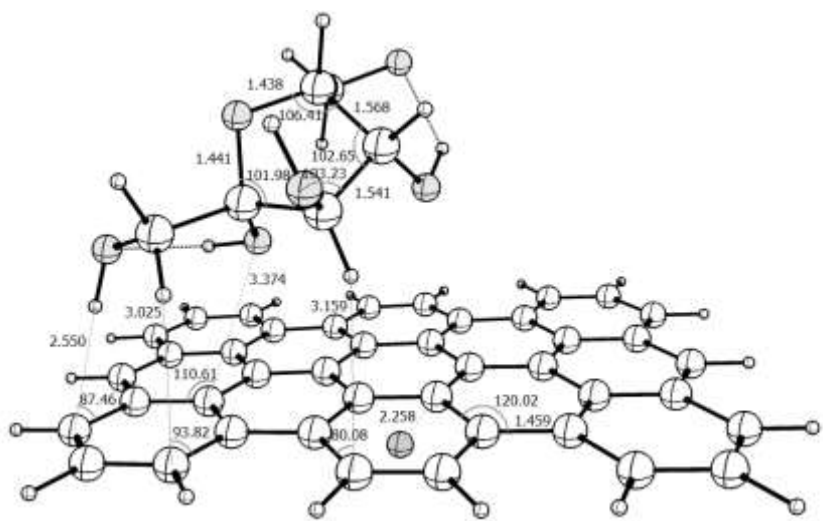

**2**

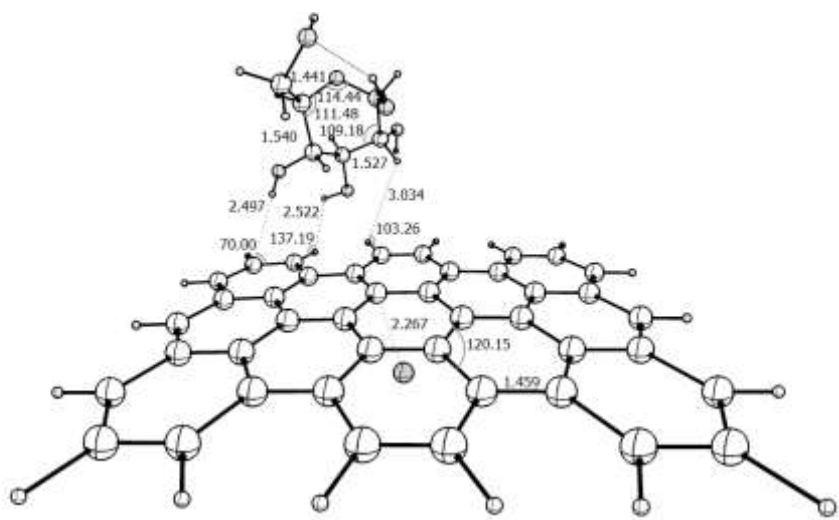

**3**

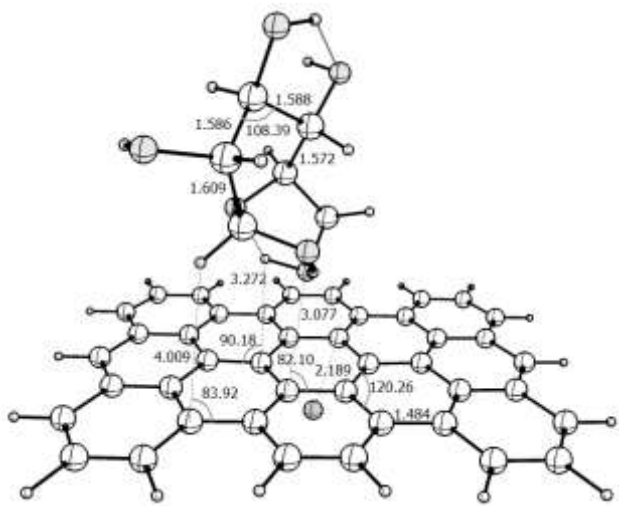

**4**

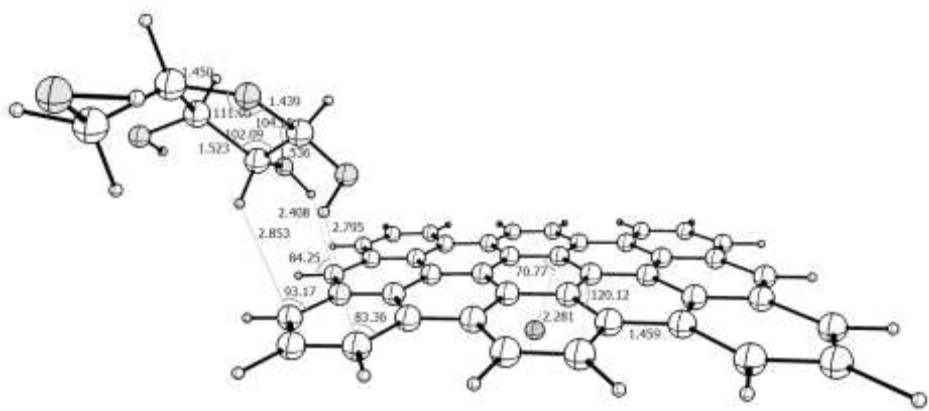

**5**

FIGURE 2.

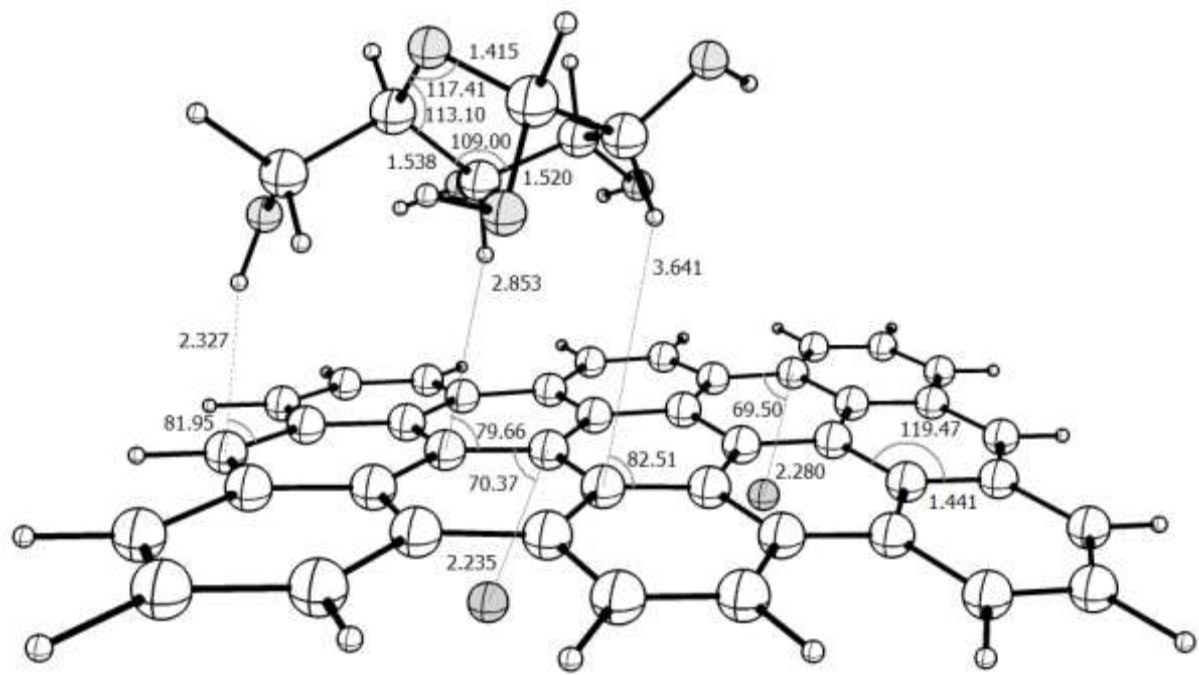

FIGURE 3.

| Sugar | No. | $\Delta E^a$ | $\Delta E^b$ |
|---|---|---|---|
| Glucose | **1** | 4.10 | 4.66 |
| Fructose | **2** | 4.07 | 4.45 |
| Galactose | **3** | 3.61 | 4.34 |
| Mannose | **4** | 4.97 | 4.93 |
| Ribose | **5** | 2.59 | 5.28 |

TABLE 1.